\documentclass[referee]{aa}
\usepackage{natbib}
\bibpunct{(}{)}{;}{a}{}{,}

\usepackage[dvips]{graphicx}
\usepackage{times}
\newcommand\ignore[1]{} 
\begin{document}
\title{Color-Induced Displacement double stars in SDSS}

\author{ D.~Pourbaix\inst{1,2}\fnmsep\thanks{Research Associate, F.N.R.S., Belgium}\and {\v Z}.~Ivezi\'c\inst{1,3} \and G.~R.~Knapp\inst{1}\and J.~E.~Gunn\inst{1}}

\offprints{D. Pourbaix (ULB)}

\institute{Department of Astrophysical Sciences, Princeton University,
Princeton, NJ 08544-1001, USA; pourbaix,ivezic,gk,jeg@astro.princeton.edu
\and 
Institut d'Astronomie et d'Astrophysique, Universit\'e
Libre de Bruxelles, CP. 226, Boulevard du Triomphe, B-1050
Bruxelles, Belgium; pourbaix@astro.ulb.ac.be
\and
H.N. Russell Fellow, on leave from the University of Washington
}
\date{Received date; accepted date} 
 
\authorrunning{Pourbaix et al.}
\titlerunning{CID double stars in SDSS}

\abstract{
We report the first successful application of the astrometric
color-induced displacement technique (CID, the displacement of
the photocenter between different bandpasses due to a varying
contribution of differently colored components to the total
light), originally proposed by \citet{Wielen-1996} for discovering
unresolved binary stars. Using the Sloan Digital Sky Survey (SDSS)
Data Release 1 with $\sim 2.5\,10^6$ stars brighter than 21$^m$ in the $u$
and $g$ bands, we select 419 candidate binary stars with CID greater
than 0.5 arcsec. The SDSS colors of the majority of these candidates
are consistent with binary systems including a white dwarf and any
main sequence star with spectral type later than $\sim$K7. The
astrometric CID method discussed here is complementary to the
photometric selection of binary stars in SDSS discussed by
\citet{Smolcic-2004:a}, but there is considerable overlap
(15\%) between the two samples of selected candidates. This overlap
testifies both to the physical soundness of both methods,
as well as to the astrometric and photometric quality of SDSS data.
}
\maketitle

\section{Introduction}

It is believed that 50\% of all stars belong to multiple systems \citep{Heintz-1969:a}.  Nevertheless, being aware that a specific star is a binary is always useful because either one throws it out of the sample or updates the model to describe it and begins some follow-up observations!  Whether one deals with stellar evolution or galactic dynamics, binaries always receive some special considerations.  So, it is important to be able to detect the binary nature of a star at an early stage of an investigation by any possible means.  

Besides spectroscopy, photometry, and interferometry, astrometry coupled to photometry has lately emerged as a way of revealing the binary nature of a source \citep{Wielen-1996}.  That method relies upon either a change in the position of the source as its brightness varies (Variability-Induced Movers, VIM) or a photometric-band dependence of the position (Color-Induced Displacement, CID).  Whereas VIM requires several observations along the brightness variation cycle and only works when at least one component is variable, it only takes one image in each band to identify a CID and can be done for non-variable stars.  Despite the number of multi-band photometric surveys, none so far has carried sufficiently accurate astrometry in at least two distinct bands.  This lack of observations has been recently alleviated by the Sloan Digital Sky Survey.

The Sloan Digital Sky Survey \citep[SDSS;][ and references therein]{York-2000:a,Abazajian-2003:a} is revolutionizing stellar astronomy by providing homogeneous and deep ($r < 22.5$) photometry in five passbands \citep[$u$, $g$, $r$, $i$, and $z$;][]{Fukugita-1996:a,Gunn-1998:a,Hogg-2001:a,Smith-2002:b} accurate to 0.02 mag \citep{Ivezic-2003:a}. Ultimately, up to 10,000 deg$^2$ of sky in the Northern Galactic Cap will be surveyed. The survey sky coverage will result in photometric measurements for over 100 million stars and a similar number of galaxies. Astrometric positions are accurate to better than 0.1 arcsec per coordinate (rms) for point sources with $r<20.5^m$ \citep{Pier-2003:a}, and the morphological information from the images allows robust star-galaxy separation to $r \sim$ 21.5$^m$ \citep{Lupton-2003:a}.

Using the SDSS data, we report on the first successful identification of Color-Induced Displacement (hereafter CID) binaries.  The underlying ideas of that method are given in Sect.~\ref{sect:cid}.   Sect.~\ref{sect:simu} describes the simulation that allowed us to optimize the screening of the data described in Sect.~\ref{sect:data}.  In Sect.~\ref{sect:results}, we present our results and compare them with those of \citet{Smolcic-2004:a}, who have recently used color selection to identify a stellar locus made of white dwarf+M dwarf binaries.
.

\section{Color Induced Displacement}\label{sect:cid}

\begin{figure}
\resizebox{\hsize}{!}{\includegraphics{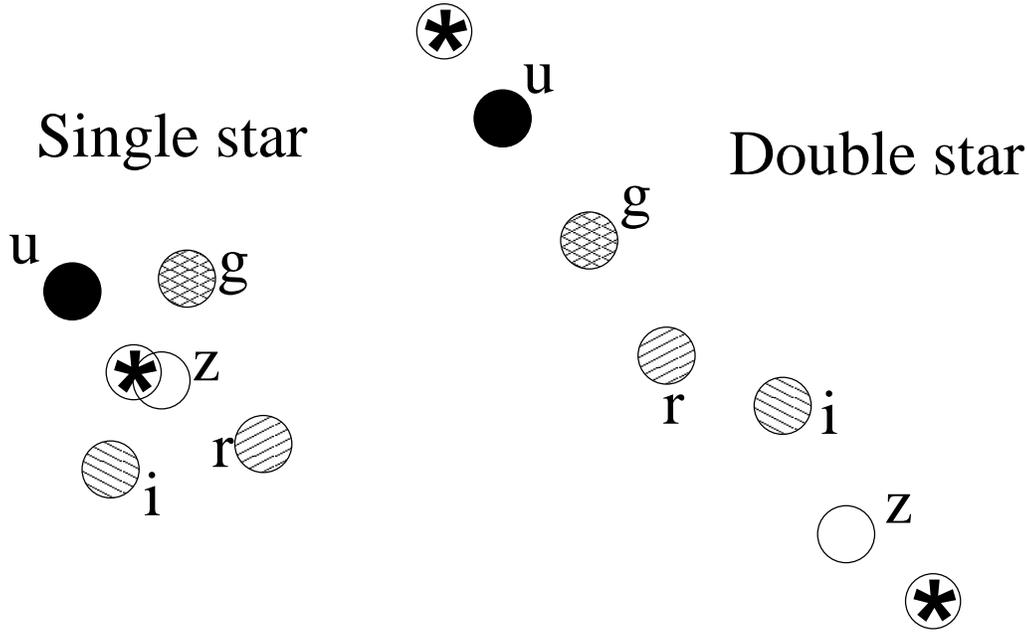}}
\caption[]{\label{Fig:cid}Position of the photocenter in the different SDSS bands.  For double stars, the positions are aligned with the two stars and their order follows their central wavelength.  Measurement error prevents the positions from being perfectly superposed/aligned for a single/double star.  The true position of the star(s) is represented as a five-branch `star'}
\end{figure}

For any unresolved double star and any photometric filter, the position of the photocenter lies between the two components.  If the two components have different colors, the position of the photocenter will change with the adopted filter.  The color-induced displacement is the change of the position of an unresolved binary depending on the adopted filter (Fig.~\ref{Fig:cid}).

Though \citet{Wielen-1996} suggested that variability induced motion and color induced motion could reveal the binary nature of an unresolved star, only the first approach has been applied to date, in the framework of Hipparcos \citep{Wielen-1996,Hipparcos,Pourbaix-2003:b}.  However, during the preparation of the Tycho-2 catalogue \citep{Hog-2000:a}, Tycho and 2MASS \citep{Skrutskie-1997:a} positions were compared.  Source duplicity came up as a satisfactory explanation for most of the discrepant positions (S.~Urban \& V. Makarov, priv. comm.).

In the case of SDSS, the position of any object together with its magnitude are measured in five bands \citep{Pier-2003:a}.  Even after the chromaticity effects have been accounted for, the five positions of a single star do not superpose exactly owing to measurement error (left panel of Fig.~\ref{Fig:cid}).  For double stars, $u$ and $z$ photometric bands will yield the two positions which are the most separated because of the largest central wavelength difference between these two filters (right panel).  

The position of the photocenter follows the peak of the efficiency of the filter.  Therefore, the photocenters are not only aligned but also ordered by the filter effective wavelengths.  The $r$ band therefore plays a central role even if the $r$ photocenter does not necessarily lie right at the middle of the $u$ and $z$ photocenters.  Instead of requiring that the $u$-, $r$-, and $z$-photocenters are aligned and well ordered, one can require that the angle measured from the $r$-position between $u$ and $z$ is 180\degr.

From now on, we will refer to the $(u,z)$ angle as the angle measured from the $r$-position between $u$ and $z$.  The distance between $u$ and $z$ (noted $\|(u,z)\|$) will refer to the angular separation between the position of the photocenter measured in the $u$ and $z$ band, respectively.  

We do not use the $g$- and $i$-positions because the scatter on the $g$- and $i$-photocenter is not significantly smaller than on the $u$ and $z$ positions, thus leading to a lower signal-to-noise ratio (S/N) as far as $\|(g,i)\|$ is concerned.
 
\section{Simulations}\label{sect:simu}
In order to develop expectations for the relevant parameter space, we first carry out some simulations with noise properties consistent with the actual data, to estimate a lower bound of the separation we can expect to notice.  According to \cite{Pier-2003:a}, the standard deviation on the relative position, i.e. band-to-band, for objects brighter than $r\sim 20$~mag is 31 mas in $u$ and 27 mas in $z$ but the authors did not quote any correlation between the residuals in those two bands.

A parent population of $\sim 2.4\,10^6$ stars with $u<21$ and $g<21$ and good photometry (see next section for details) was therefore used to update the precision of the residuals and to derive the correlation between them.  The residuals in right ascension in $u$ and $z$ have a standard deviation of respectively 36 and 20 mas, with a correlation of 0.16.  In declination, these precisions are 39 and 22 mas for $u$ and $z$ with a correlation of 0.18.

\begin{figure}
\resizebox{\hsize}{!}{\includegraphics{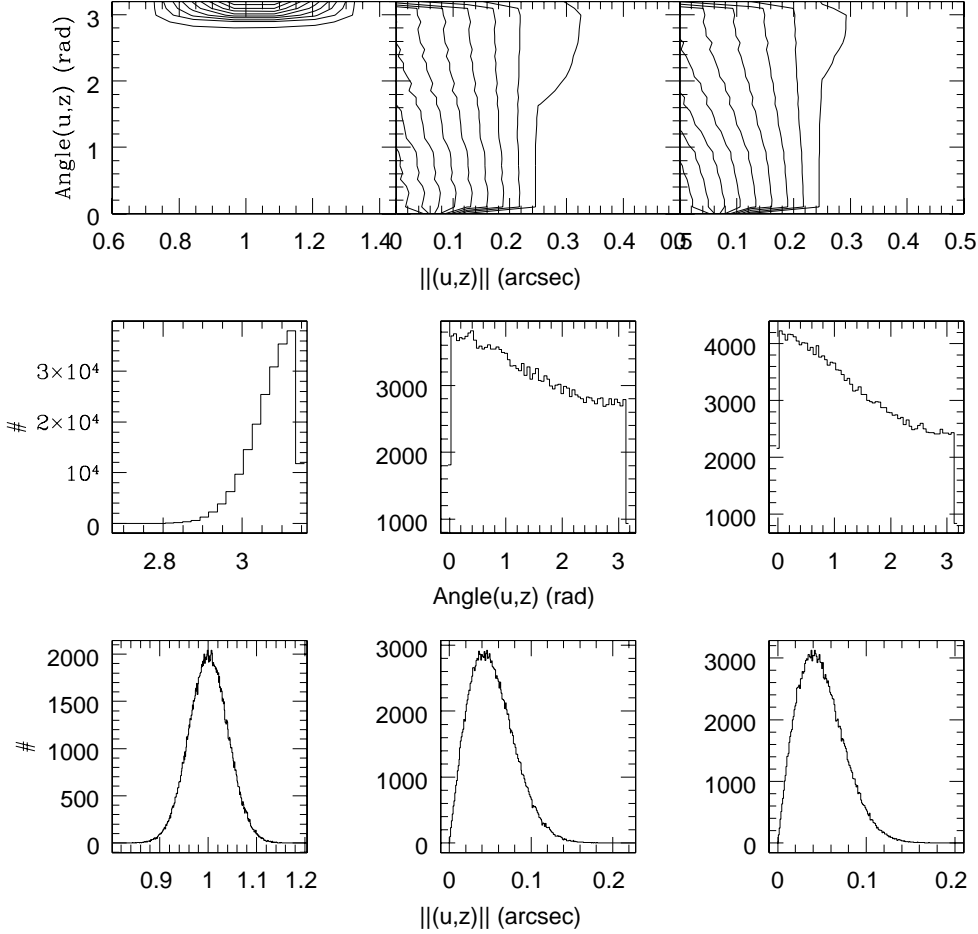}}
\caption[]{\label{Fig:sim1}Model distribution of the angles and separations between the photocenters in $u$ and $z$ assuming a true separation of 1\arcsec (left column), 20 mas (central column) and 0 mas (right column).  Each line in the contour (top panels) represents a linear 10\% increment.}
\end{figure}

For the simulation, the origin of the offset is placed at the position in the $r$ band and is assumed to lie in the middle of the segment joining the true photocenters in $u$ and $z$.  The distributions of the angle and separation between the $u$- and $z$-positions are derived from 200\,000 model positions generated for both photocenters using the above standard deviations and correlations.  Such distributions for separations of 1\arcsec, 20 mas and 0 mas are plotted in Fig.~\ref{Fig:sim1}.  The spread in the angle increases as the true separation goes to zero.  Using a 99.9\% confidence level Kolmogorov-Smirnov test, we reject the hypothesis that the distributions of $(u,z)$ for single stars and binaries with separations below 10 mas are different.  It is worth noting that in the latter case, the correlation between the $u$- and $z$-positions prevents the angles from being uniformly distributed over 0-$\pi$.

Though the percentage of binaries is usually thought to be quite high (ranging from 85\% among OB stars \citep{Heintz-1969:a} down to 30\% for M stars \citep{Marchal-2003:a} and even 10\% for F7-K systems with periods less than 10 years \citep{Halbwachs-2003:a}), very few binaries actually induce a noticeable shift of the photocenter.  While the color induced displacement is larger the more different the colors, too large a magnitude difference causes the fainter component to be undetectable.  Thus only a small fraction of true binaries will show detectable CID.  Coupled to the spread of the angle due to measurement error, this means that one should not expect a large deviation of the distribution with respect to that of a pure single star population, especially if one looks at the whole SDSS sample.

\section{Data}\label{sect:data}

In this study, we use the public SDSS data release 1 (DR1) which contains a few million stars.  Because errors are larger towards fainter magnitudes, we selected stars to be bright enough to have good astrometry ($u,g<21$), thus reducing the number of stars to $\sim 2.5\,10^6$.  We further impose
\begin{enumerate}
\item the condition that the displacement between the $u$- and $z$-positions is larger than 0.2\arcsec.  This criterion is expressed as:
\[
\sqrt{(\Delta\alpha_u-\Delta\alpha_z)^2+(\Delta\delta_u-\Delta\delta_z)^2}\ge 0.2
\] 
where the four offset quantities are readily available in the SDSS database;
\item the precision on the magnitudes is better than 0.1 mag in $u$ and $r$ and better than 0.05 mag for $g$, $i$ and $z$.
\end{enumerate}
 Because astrometry and photometry are strongly tied together, bad photometry is likely to show up as poor astrometry anyway so these constraints on the photometric precision are actually safeguards for the astrometry as well.  These two criteria yield a sample of only 19\,982 entries, $\sim 0.8$\% of the previous sample.

\begin{figure}
\resizebox{\hsize}{!}{\includegraphics{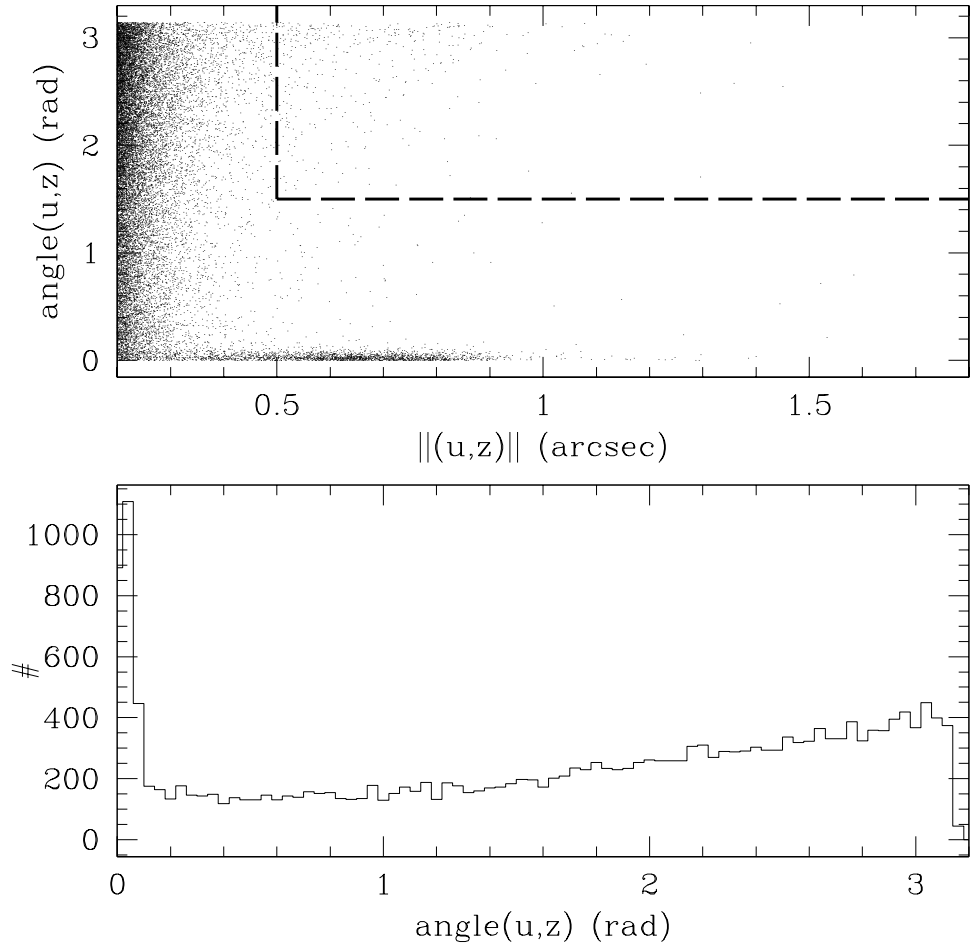}}
\caption[]{\label{Fig:angldist}Observed distribution of the angles and separations between the photocenters in $u$ and $z$.  The upper-right corner delimited by the dashed lines corresponds to our revised selection criterion.}
\end{figure}

The angles versus the separations between the $u$- and $z$-photocenters the distribution of the latter are plotted in Fig.~\ref{Fig:angldist}.  With respect to the simulations from Sect.~\ref{sect:simu}, there is a strong excess of points with a large displacement but with a 0 angle.  This very striking feature is due to asteroids.  Both the displacement and colors of all these 2\,180 objects are consistent with the asteroids already identified in SDSS \citep{Ivezic-2002:a,Juric-2002:a}.  In the case of asteroids, the position of the photocenter changes from $u$ to $z$ because of the genuine displacement of the object on the sky in between the two exposures.  Because of the way the $u$ and $z$ CCDs are located on the detector (the scanning order is $rizug$), the positions in $u$ and $z$ are on the same side with respect to the $r$-positions, thus yielding a null angle.

In the single star simulation, the distribution of the angle between the $u$ and $z$-photocenters shows a continuous decrease towards $\pi$.  However the lower panel of Fig.~\ref{Fig:angldist} reveals that, once out of the asteroid region of the distribution, the number of large angles actually increases, which is consistent with the presence of binaries.

Even though the distribution shows an increase from 0.2 rad up to $\pi$ and the simulation indicates that no single star is likely to cause a displacement larger than 0.3\arcsec, we conservatively adopt 1.5 rad and 0.5\arcsec\ for the lower bounds of the angle and separation respectively.  This final cut leads to 419 candidate binaries.  There is no noticeable clustering of these objects in chip coordinates nor in $(\alpha, \delta)$ which rules out the possibility of an instrumental (e.g. corrupted CCD row or column) or observational (e.g. unnoticed bad seeing) problem which would have caused the displacement.

\section{Results}\label{sect:results}

\subsection{WD+MD bridge}\label{sect:wdMd}

If the components of a binary have different colors, the resulting colors of the system do not all match single star ones.  Whereas most combinations of stellar colors remain consistent with the stellar locus, any departure from the latter can be interpreted as the signature of the duplicity of the source.  \citet{Smolcic-2004:a} used that color selection to identify a second stellar locus in the SDSS data, namely a bridge in the $u-g$ vs $g-r$ color-color diagram between white and M dwarfs. 

\begin{figure}
\resizebox{\hsize}{!}{\includegraphics{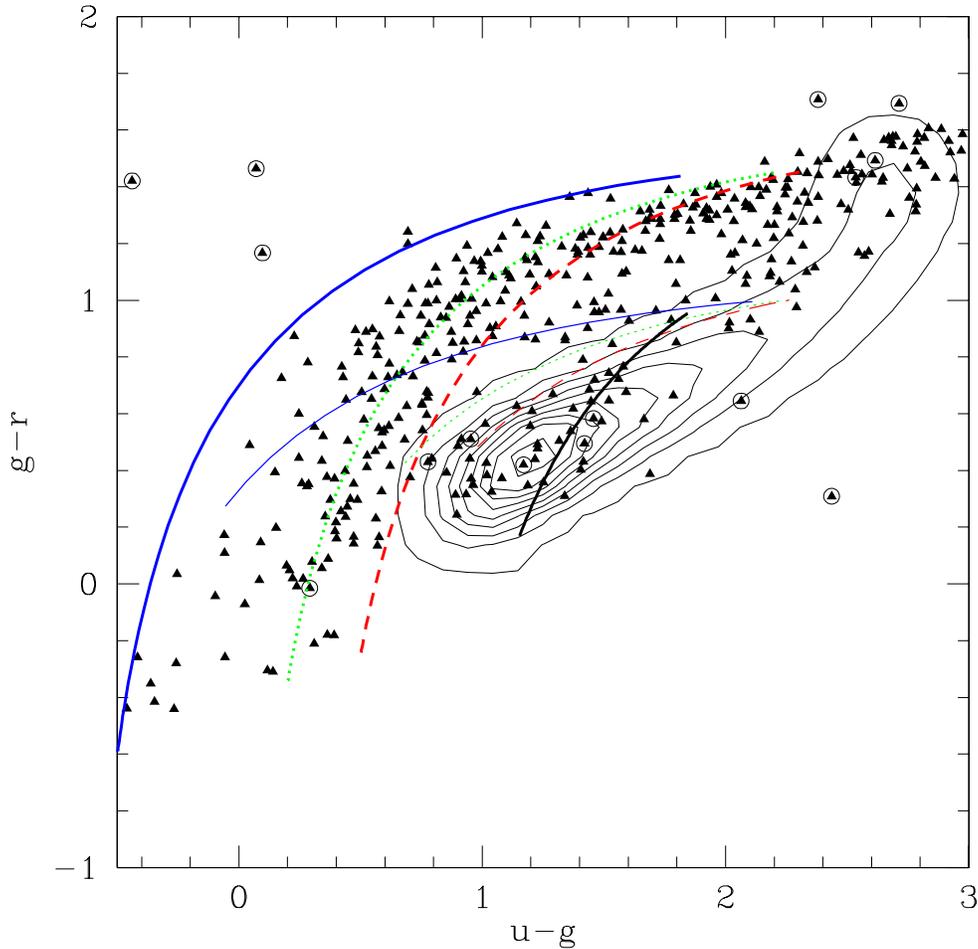}}
\caption[]{\label{Fig:ccSDSSuggr}Color-color diagram of the putative binaries (triangles) superposed over the original parent population of 284\,503 stars (contours). The thick/thin lines represent systems with a M dwarf/K7V component.  The short thick line close to the center corresponds to A0V+K5III systems.  Triangles with a circle around have weird colors that could be the cause of the displacement.} 
\end{figure}

Our 419 candidate binaries are displayed in the same color-color diagram in Fig.~\ref{Fig:ccSDSSuggr}.  Although the two techniques are rather orthogonal, our purely astrometric approach reproduces the essence of the color-based results after \citet{Smolcic-2004:a}.  Such a result was expected since those two groups of stars have the largest color difference but are of similar absolute magnitudes and are therefore the most likely to yield a noticeable displacement between the $u$- and $z$-photocenters.  Only 61 objects match the criteria of brightness and colors imposed by Smol\v{c}i\'c et al, i.e. 7\% of their bridge stars match our astrometric criteria.

In order to explain the bottom-left part of the diagram, one cannot rely on the assumption of a unique color for all the white dwarfs, especially in $u-g$ and $g-r$.  Following \citet{Harris-2003:a} we instead adopt three different model WDs to cover the range of $u-g$, yet keeping $r-i=i-z=0$.  The coordinates of the three WDs in the $(u-g,g-r)$ plane are respectively $(-0.5,-0.6)$, $(0.2,-0.35)$, $(0.5,-0.25)$.  The WD colors adopted by \citet{Smolcic-2004:a} lies between our second and third models.  The flux ratio between the white dwarf and the M dwarf is assumed to range between 0.01 and 100.0.  The three resulting tracks are plotted as thick lines in Fig.~\ref{Fig:ccSDSSuggr}.

Since at least one of our extended bridges passes over the quasar region of the $(u-g,g-r)$ diagram, could it be that some of our candidates binaries are actually QSOs?  From the $i-z$ of these points (Fig.~\ref{Fig:ccSDSSugiz}), one can conclude that it is seldom (if ever) the case (according to \citet{Richards-2001:a}, QSOs have $i-z\sim 0$). All the questionable points are hence consistent with a white dwarf + M dwarf pair.

\subsection{Additional models}

Owing to the narrow convergence of the three tracks at the M dwarf end, only 50\% of the points are bounded by the previous models. What other combinations of stars would produce system colors consistent with the data?  As we have already stated, the colors have to be rather different and yet the magnitudes to be rather similar.  Assuming a WD companion, the bluer the main sequence component, the larger the magnitude difference and the more similar the colors. 

If the M dwarf is replaced with, say, a K7 main-sequence star (assuming 2.32, 1.01, 0.32, and 0.15 for $u-g$, $g-r$, $r-i$, and $i-z$ respectively based on the list after \citet{Gunn-1983:a} convolved with the SDSS filters \citep{Fukugita-1996:a}), the flux ratio in the $r$ band becomes a bit more constrained: $2.5F_{r,K}\le F_{r,WD}<250F_{r,K}$.  The different tracks leading to the same three model white dwarfs as in Sect.~\ref{sect:wdMd} are plotted as thin lines in Fig.~\ref{Fig:ccSDSSuggr}.  Because of the flux ratio constraint, these tracks do not go down to the WD region but that part of the diagram is already covered by the WD+MD model anyway.  

\begin{figure}
\resizebox{0.49\hsize}{!}{\includegraphics{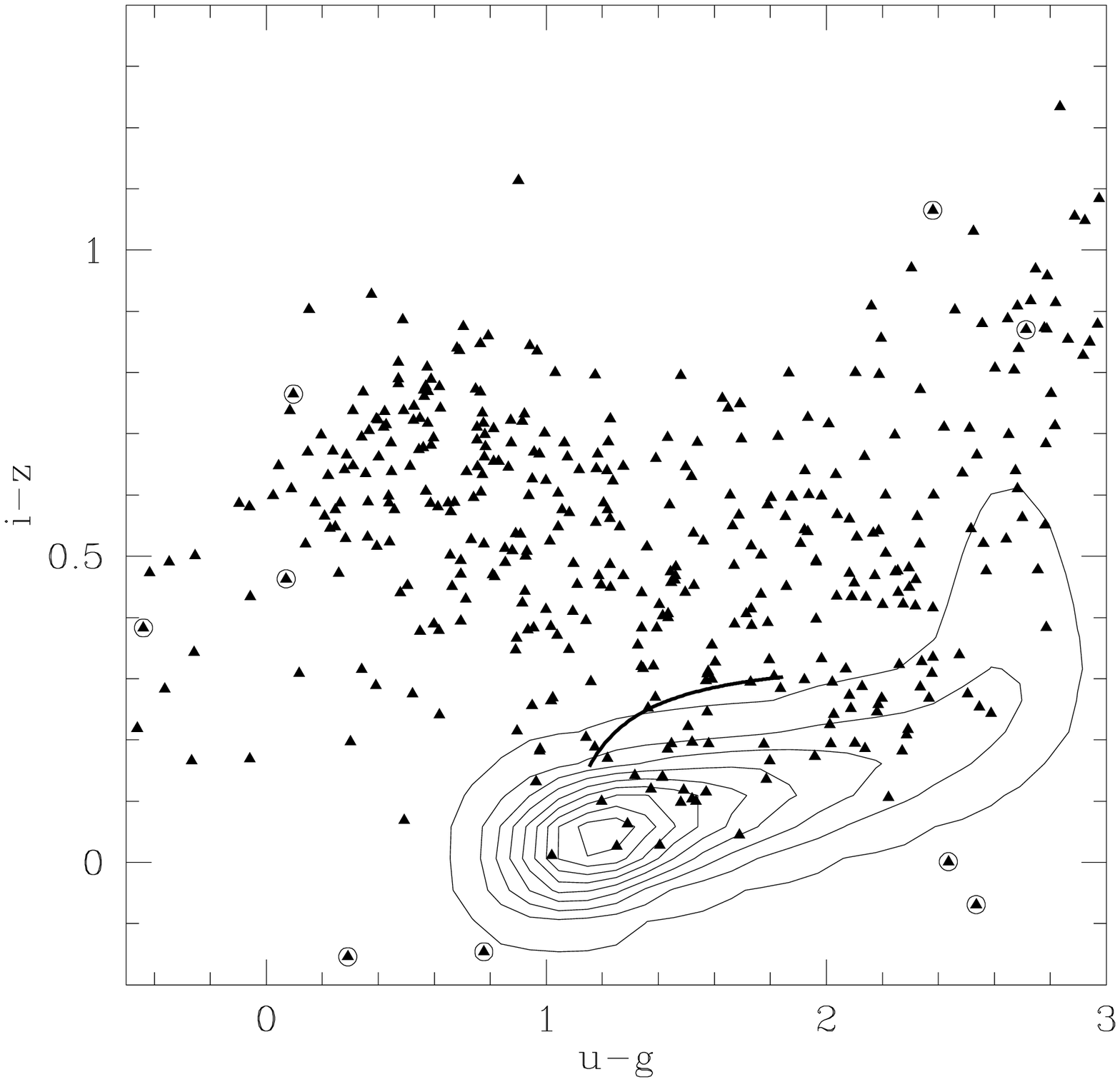}}\hfill
\resizebox{0.49\hsize}{!}{\includegraphics{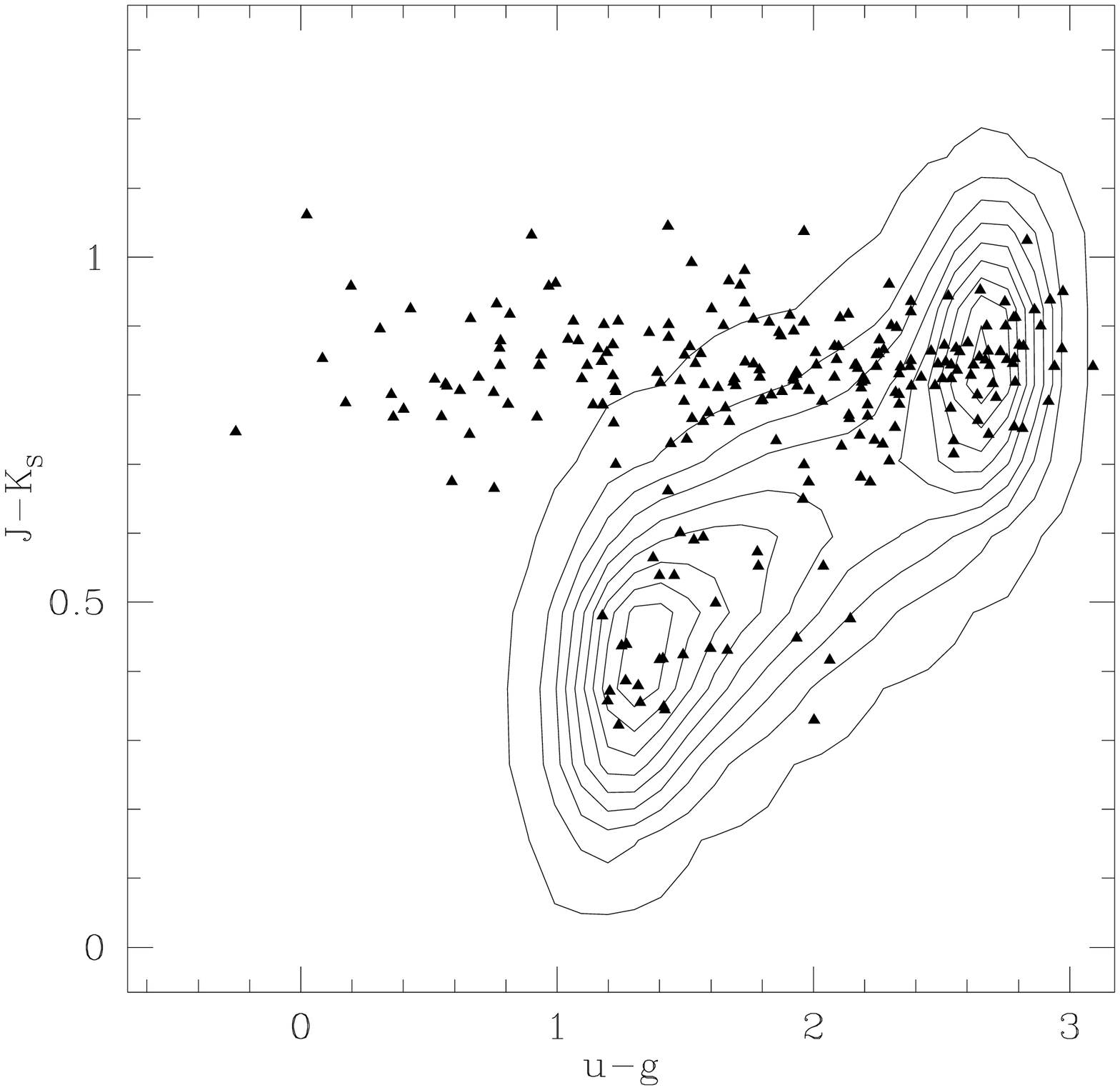}}
\caption[]{\label{Fig:ccSDSSugiz}{\bf Left panel} Alternative color-color diagram.  See caption of Fig,~\ref{Fig:ccSDSSuggr} for details.  {\bf Right panel} SDSS-2MASS color-color diagram of 241 SDSS CID binaries matched in the 2MASS database}
\end{figure}

Almost 90\% of the candidate binaries can thus be explained with a model involving a WD and any main-sequence star redder than K7.  What about the remaining 10\% whose $(u-g,g-r)$ cannot that easily be explained with any of the previous scenarios?  Even though several of these points are located around $(1.1,0.4)$ in the $(u-g,g-r)$ diagram, i.e. they are consistent with the peak of the stellar locus and hence with a possible contamination by weird single stars, that region is totally depleted in the $(u-g,i-z)$ diagram (Fig.~\ref{Fig:ccSDSSugiz}).

Besides the binaries with a WD component, what other systems could exhibit large displacements between the $u$ and $z$ filters?  A A0 main-sequence and a giant K5 have rather similar magnitudes, yet rather different colors (almost as much as the WD+MD pairs).  The flux ratio in the $r$ band between the two components is constrained to the range 0.63--6.3.  For the K5 giant (resp. A0V), we adopt the colors 3.40 (1.01), 1.35 (-0.21), 0.56 (-0.14), and 0.33 (-0.11) for $u-g$, $g-r$, $r-i$, and $i-z$ respectively \citep{Gunn-1983:a,Fukugita-1996:a}.  The corresponding track is displayed in both Fig.~\ref{Fig:ccSDSSuggr} and \ref{Fig:ccSDSSugiz} (left panel) as a thick line.  It is worth noting that this track is consistent with the stellar locus in both diagrams thus confirming that the CID binaries would not all be identified as outliers in a color-color diagram.

Even though 2MASS does not go as faint as SDSS, the overlap between the two surveys \citep{Finlator-2000:a} is nevertheless large enough to obtain combined color-color diagrams for 241 of our putative double stars ($\sim 60$\%).  Among the combinations of filters, the $u-g$ vs $J-K_S$ exhibits the largest departure of these binaries from the stellar locus (right panel of Fig.~\ref{Fig:ccSDSSugiz}).  Instead of a bridge as in Fig.~\ref{Fig:ccSDSSuggr}, the binary locus appears as a narrow horizontal band in th combined 2MASS SDSS diagram.  Unlike the SDSS colors, the 2MASS color allows us to rule out the possibility for the M stars to be giant rather than dwarf.  Note also that there are still CID binaries whose colors are very consistent with the main stellar locus even when 2MASS bands are used as well. 

\subsection{Contamination} 
About 15 stars have weird colors (triangles with a circle around in Figs.~\ref{Fig:ccSDSSuggr} and \ref{Fig:ccSDSSugiz}), especially in $i-z$, which is below -0.5 for six of them.  In such cases, the noticed displacement would be caused by the wrong astrometric transformation induced by the unusual colors rather than by a true displacement of the photocenter. However spectra available for some of these troublesome points confirm the duplicity.  So, in the worst case, the contamination rate does not exceed 3\%.  

\subsection{Angular separation}

\begin{figure*}
\resizebox{0.3\hsize}{!}{\includegraphics{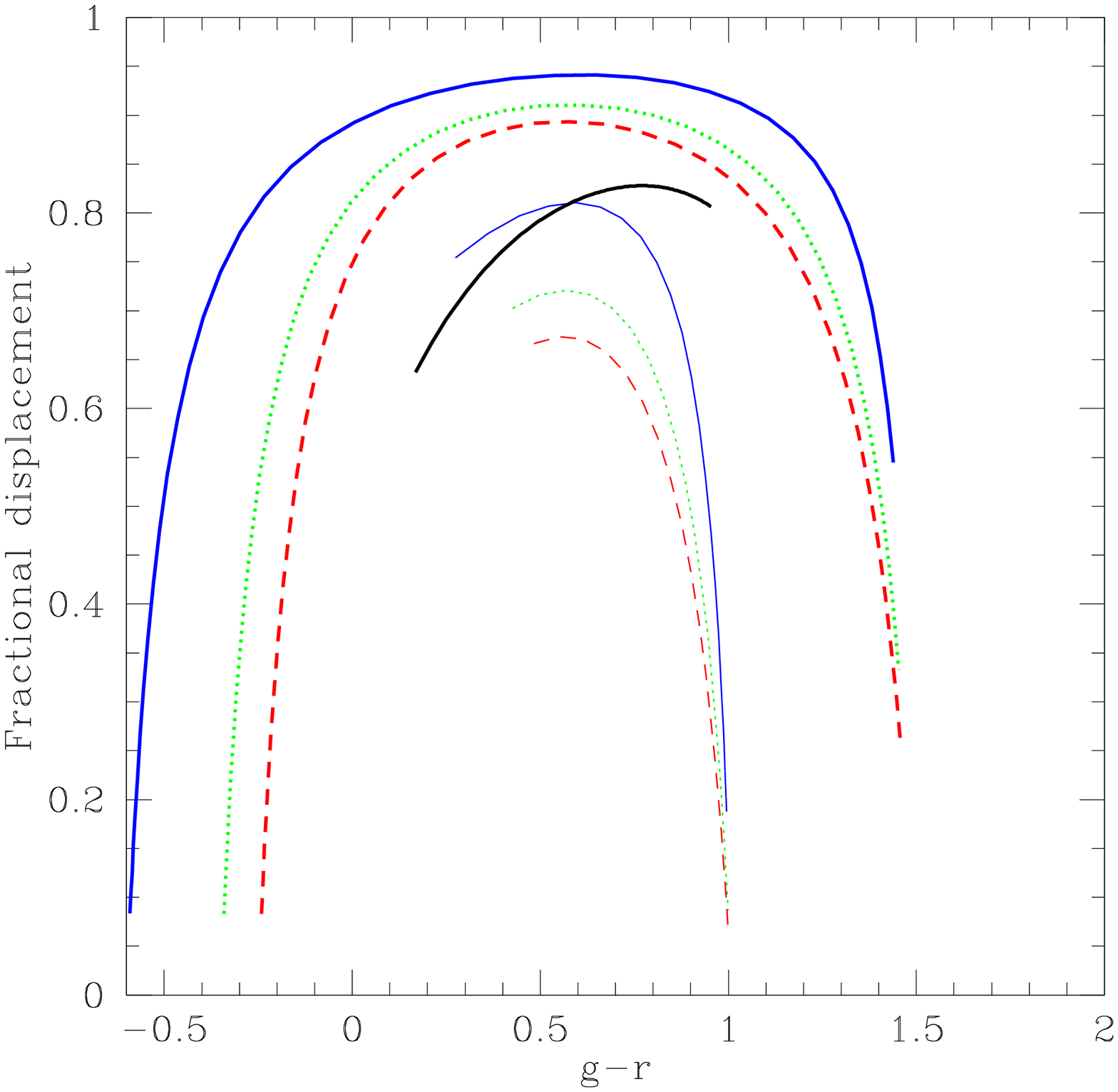}}
\resizebox{0.3\hsize}{!}{\includegraphics{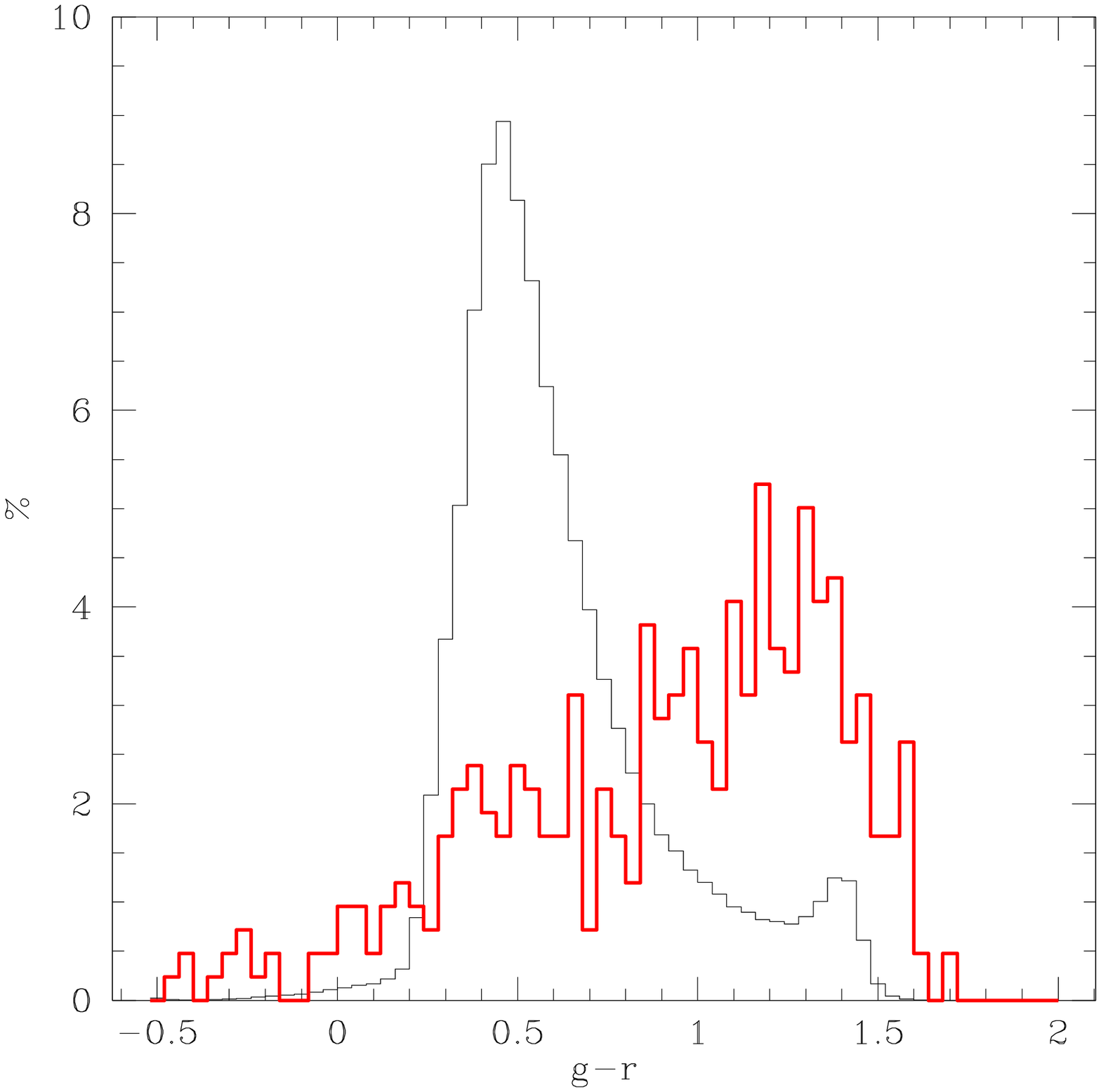}}
\resizebox{0.3\hsize}{!}{\includegraphics{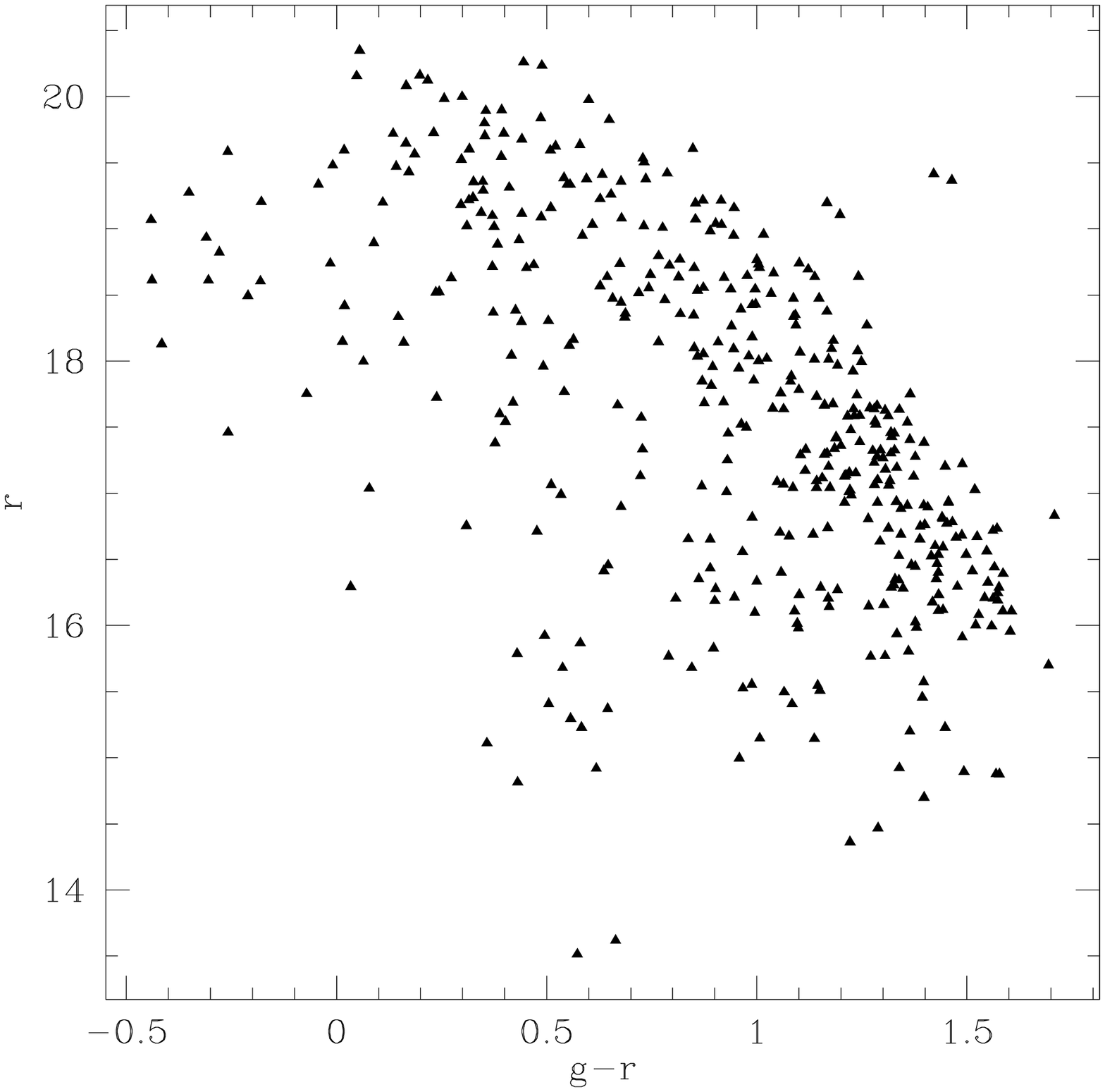}}
\caption[]{\label{Fig:delta}Left panel: Ratio of $\|(u,z)\|$ over the actual angular separation of the two components.  The thick (resp. thin) lines represent systems with a M dwarf (resp. K7V) component.  The short thick line in the upper right corresponds to A0V+K5III systems. Central panel: Distribution of the $g-r$ color of the parent population (thin line) and the binaries (thick line).  Right panel: Color-magnitude of the putative binaries}
\end{figure*}

The likelihood of detecting a CID binary clearly depends on the difference in colors of the components but it also depends on the actual angular separation of the two stars.  From the only criterion adopted so far, namely $\|(u,z)\|>0.5$\arcsec, what can we infer about that angular separation?  The ratio of $\|(u,z)\|$ to that separation is plotted in the left panel of Fig.~\ref{Fig:delta}.

All the white dwarf systems show a maximum displacement above $g-r=0.5$, corresponding to 60\%, so any separation larger than 0.85\arcsec\ fulfills the criterion.  That is consistent with the distribution plotted in the central panel of Fig~\ref{Fig:delta}.  It shows that whereas the parent distribution peaks below $g-r=0.5$, the binary distribution peaks well above that value.  

Whereas the fractional displacement starts decreasing above $g-r=0.7$, the binary distribution keeps growing up to $g-r\sim 1.3$ where the displacement has already decreased to 70\%.  This means that the decrease of the fractional displacement is compensated by the actual angular separation of the components.  Assuming a constant linear separation, these systems should thus be closer.  A color-magnitude diagram (right panel of Fig.~\ref{Fig:delta}) gives credit to that explanation.  Indeed, the redder, the brighter while the absolute magnitude goes up thus meaning that the reddest objects are on average closer to us.

\section{Conclusions}
The color induced displacement method described by \citet{Wielen-1996} as a way of detecting binaries has been successfully applied to the first public release of the SDSS data.  We identify about 400 systems whose changes in position are essentially consistent with a white dwarf coupled to a lower end (later than $\sim$K7) main-sequence star.  We therefore expect $\sim2\,000$ CID binaries at the completion of the SDSS observation campaign.

This identification of binaries is an independent confirmation of the color based results of \citet{Smolcic-2004:a}.  However, whereas they had a lower bound on $g-r$ of 0.3, the astrometric criterion allows us to identify candidate binaries down to $g-r=-0.4$.  On the other hand, color selection they utilized is more sensitive to binaries with angular separations smaller than the sensitivity of the CID method.

Though the approach has proven to give results, its efficiency is extremely low.  Whereas \citet{Marchal-2003:a} quote at least 30\%\ of binaries among M stars (the percentage grows with the mass of the star along the main-sequence), only 0.02\%\ are detected through their CID effect.  It is noteworthy that with such a low fraction, the CID binaries do not affect the overall SDSS astrometric precision.

Because of their much better astrometric precision (typically a few $\mu$as), space-based astrometry missions like SIM and Gaia will eventually supersede the SDSS results presented here.  According to a Gaia preparatory study \citep{Arenou-2001:a}, the latter could, for instance, detect a M0 companion to a G0 dwarf star at a $3\sigma$ level at a separation as low as 2.3 mas.  In terms of separations, this is $\sim 200$ times better than the sensitivity of the CID method applied to the SDSS data.

\begin{acknowledgements}
This work is partly supported by NASA grant NAG5-11094 to Princeton University. ZI and GK thanks Princeton University for generous support of their work.  Funding for the creation and distribution of the SDSS Archive has been provided by the Alfred P. Sloan Foundation, the Participating Institutions, the National Aeronautics and Space Administration, the National Science Foundation, the U.S. Department of Energy, the Japanese Monbukagakusho, and the Max Planck Society.  The SDSS web site is http://www.sdss,org/.
\end{acknowledgements}

\bibliographystyle{aa} 
\bibliography{articles,books}

\begin{thebibliography}{25}
\expandafter\ifx\csname natexlab\endcsname\relax\def\natexlab#1{#1}\fi

\bibitem[{{Abazajian} {et~al.}(2003){Abazajian}, {Adelman-McCarthy},
  {Ag{\"u}eros}, {Allam}, {Anderson}, {Annis}, {Bahcall}, {Baldry}, {Bastian},
  {Berlind}, {Bernardi}, {Blanton}, {Blythe}, {Bochanski}, {Boroski},
  {Brewington}, {Briggs}, {Brinkmann}, {Brunner}, {Budav{\' a}ri}, {Carey},
  {Carr}, {Castander}, {Chiu}, {Collinge}, {Connolly}, {Covey}, {Csabai},
  {Dalcanton}, {Dodelson}, {Doi}, {Dong}, {Eisenstein}, {Evans}, {Fan},
  {Feldman}, {Finkbeiner}, {Friedman}, {Frieman}, {Fukugita}, {Gal},
  {Gillespie}, {Glazebrook}, {Gonzalez}, {Gray}, {Grebel}, {Grodnicki}, {Gunn},
  {Gurbani}, {Hall}, {Hao}, {Harbeck}, {Harris}, {Harris}, {Harvanek},
  {Hawley}, {Heckman}, {Helmboldt}, {Hendry}, {Hennessy}, {Hindsley}, {Hogg},
  {Holmgren}, {Holtzman}, {Homer}, {Hui}, {Ichikawa}, {Ichikawa}, {Inkmann},
  {Ivezi{\' c}}, {Jester}, {Johnston}, {Jordan}, {Jordan}, {Jorgensen},
  {Juri{\' c}}, {Kauffmann}, {Kent}, {Kleinman}, {Knapp}, {Kniazev}, {Kron},
  {Krzesi{\' n}ski}, {Kunszt}, {Kuropatkin}, {Lamb}, {Lampeitl}, {Laubscher},
  {Lee}, {Leger}, {Li}, {Lidz}, {Lin}, {Loh}, {Long}, {Loveday}, {Lupton},
  {Malik}, {Margon}, {McGehee}, {McKay}, {Meiksin}, {Miknaitis}, {Moorthy},
  {Munn}, {Murphy}, {Nakajima}, {Narayanan}, {Nash}, {Neilsen}, {Newberg},
  {Newman}, {Nichol}, {Nicinski}, {Nieto-Santisteban}, {Nitta}, {Odenkirchen},
  {Okamura}, {Ostriker}, {Owen}, {Padmanabhan}, {Peoples}, {Pier}, {Pindor},
  {Pope}, {Quinn}, {Rafikov}, {Raymond}, {Richards}, {Richmond}, {Rix},
  {Rockosi}, {Schaye}, {Schlegel}, {Schneider}, {Schroeder}, {Scranton},
  {Sekiguchi}, {Seljak}, {Sergey}, {Sesar}, {Sheldon}, {Shimasaku}, {Siegmund},
  {Silvestri}, {Sinisgalli}, {Sirko}, {Smith}, {Smol{\v c}i{\' c}}, {Snedden},
  {Stebbins}, {Steinhardt}, {Stinson}, {Stoughton}, {Strateva}, {Strauss},
  {SubbaRao}, {Szalay}, {Szapudi}, {Szkody}, {Tasca}, {Tegmark}, {Thakar},
  {Tremonti}, {Tucker}, {Uomoto}, {Vanden Berk}, {Vandenberg}, {Vogeley},
  {Voges}, {Vogt}, {Walkowicz}, {Weinberg}, {West}, {White}, {Wilhite},
  {Willman}, {Xu}, {Yanny}, {Yarger}, {Yasuda}, {Yip}, {Yocum}, {York},
  {Zakamska}, {Zehavi}, {Zheng}, {Zibetti}, \& {Zucker}}]{Abazajian-2003:a}
{Abazajian}, K., {Adelman-McCarthy}, J.~K., {Ag{\"u}eros}, M.~A., {et~al.}
  2003, AJ, 126, 2081

\bibitem[{{Arenou} \& {Jordi}(2001)}]{Arenou-2001:a}
{Arenou}, F. \& {Jordi}, C. 2001, Gaia duplicity detection: photocentric
  binaries, Tech. Rep. GAIA-FA-002, Observatoire de Paris-Meudon,
  http://wwwhip.obspm.fr/gaia/dms/texts/GAIA\_duplicity.pdf

\bibitem[{{ESA}(1997)}]{Hipparcos}
{ESA}. 1997, The Hipparcos and Tycho Catalogues (ESA SP-1200)

\bibitem[{{Finlator} {et~al.}(2000){Finlator}, {Ivezi{\' c}}, {Fan}, {Strauss},
  {Knapp}, {Lupton}, {Gunn}, {Rockosi}, {Anderson}, {Csabai}, {Hennessy},
  {Hindsley}, {McKay}, {Nichol}, {Schneider}, {Smith}, {York}, \& {the SDSS
  Collaboration}}]{Finlator-2000:a}
{Finlator}, K., {Ivezi{\' c}}, {\v Z}., {Fan}, X., {et~al.} 2000, AJ, 120, 2615

\bibitem[{{Fukugita} {et~al.}(1996){Fukugita}, {Ichikawa}, {Gunn}, {Doi},
  {Shimasaku}, \& {Schneider}}]{Fukugita-1996:a}
{Fukugita}, M., {Ichikawa}, T., {Gunn}, J.~E., {et~al.} 1996, AJ, 111, 1748

\bibitem[{{Gunn} {et~al.}(1998){Gunn}, {Carr}, {Rockosi}, {Sekiguchi}, {Berry},
  {Elms}, {de Haas}, {Ivezi{\' c}}, {Knapp}, {Lupton}, {Pauls}, {Simcoe},
  {Hirsch}, {Sanford}, {Wang}, {York}, {Harris}, {Annis}, {Bartozek},
  {Boroski}, {Bakken}, {Haldeman}, {Kent}, {Holm}, {Holmgren}, {Petravick},
  {Prosapio}, {Rechenmacher}, {Doi}, {Fukugita}, {Shimasaku}, {Okada}, {Hull},
  {Siegmund}, {Mannery}, {Blouke}, {Heidtman}, {Schneider}, {Lucinio}, \&
  {Brinkman}}]{Gunn-1998:a}
{Gunn}, J.~E., {Carr}, M., {Rockosi}, C., {et~al.} 1998, AJ, 116, 3040

\bibitem[{{Gunn} \& {Stryker}(1983)}]{Gunn-1983:a}
{Gunn}, J.~E. \& {Stryker}, L.~L. 1983, ApJS, 52, 121

\bibitem[{{Halbwachs} {et~al.}(2003){Halbwachs}, {Mayor}, {Udry}, \&
  {Arenou}}]{Halbwachs-2003:a}
{Halbwachs}, J.~L., {Mayor}, M., {Udry}, S., \& {Arenou}, F. 2003, A\&A, 397,
  159

\bibitem[{{Harris} {et~al.}(2003){Harris}, {Liebert}, {Kleinman}, {Nitta},
  {Anderson}, {Knapp}, {Krzesi{\' n}ski}, {Schmidt}, {Strauss}, {Vanden Berk},
  {Eisenstein}, {Hawley}, {Margon}, {Munn}, {Silvestri}, {Smith}, {Szkody},
  {Collinge}, {Dahn}, {Fan}, {Hall}, {Schneider}, {Brinkmann}, {Burles},
  {Gunn}, {Hennessy}, {Hindsley}, {Ivezi{\' c}}, {Kent}, {Lamb}, {Lupton},
  {Nichol}, {Pier}, {Schlegel}, {SubbaRao}, {Uomoto}, {Yanny}, \&
  {York}}]{Harris-2003:a}
{Harris}, H.~C., {Liebert}, J., {Kleinman}, S.~J., {et~al.} 2003, AJ, 126, 1023

\bibitem[{{Heintz}(1969)}]{Heintz-1969:a}
{Heintz}, W.~D. 1969, JRASC, 63, 275

\bibitem[{{H\o g} {et~al.}(2000){H\o g}, {Fabricius}, {Makarov}, {Urban},
  {Corbin}, {Wycoff}, {Bastian}, {Schwekendiek}, \& {Wicenec}}]{Hog-2000:a}
{H\o g}, E., {Fabricius}, C., {Makarov}, V.~V., {et~al.} 2000, A\&A, 355, L27

\bibitem[{{Hogg} {et~al.}(2001){Hogg}, {Finkbeiner}, {Schlegel}, \&
  {Gunn}}]{Hogg-2001:a}
{Hogg}, D.~W., {Finkbeiner}, D.~P., {Schlegel}, D.~J., \& {Gunn}, J.~E. 2001,
  AJ, 122, 2129

\bibitem[{{Ivezi{\' c}} {et~al.}(2003){Ivezi{\' c}}, {Lupton}, {Anderson}, \&
  {et al.}}]{Ivezic-2003:a}
{Ivezi{\' c}}, {\v Z}., {Lupton}, R.~H., {Anderson}, S., \& {et al.} 2003, Mem.
  Soc. Ast. It., 74, 978

\bibitem[{{Ivezi{\' c}} {et~al.}(2002){Ivezi{\' c}}, {Lupton}, {Juri{\' c}},
  {Tabachnik}, {Quinn}, {Gunn}, {Knapp}, {Rockosi}, \&
  {Brinkmann}}]{Ivezic-2002:a}
{Ivezi{\' c}}, {\v Z}., {Lupton}, R.~H., {Juri{\' c}}, M., {et~al.} 2002, AJ,
  124, 2943

\bibitem[{{Juri{\' c}} {et~al.}(2002){Juri{\' c}}, {Ivezi{\' c}}, {Lupton},
  {Quinn}, {Tabachnik}, {Fan}, {Gunn}, {Hennessy}, {Knapp}, {Munn}, {Pier},
  {Rockosi}, {Schneider}, {Brinkmann}, {Csabai}, \& {Fukugita}}]{Juric-2002:a}
{Juri{\' c}}, M., {Ivezi{\' c}}, {\v Z}., {Lupton}, R.~H., {et~al.} 2002, AJ,
  124, 1776

\bibitem[{{Lupton} {et~al.}(2003){Lupton}, {Ivezi\'{c}}, {Gunn}, {Knapp},
  {Strauss}, \& {Yasuda}}]{Lupton-2003:a}
{Lupton}, R.~H., {Ivezi\'{c}}, v., {Gunn}, J.~E., {et~al.} 2003, Proc. SPIE,
  4836, 350

\bibitem[{{Marchal} {et~al.}(2003){Marchal}, {Delfosse}, {Forveille},
  {S\'egransan}, {Beuzit}, {Udry}, {Perrier}, {Mayor}, \&
  {Halbwachs}}]{Marchal-2003:a}
{Marchal}, L., {Delfosse}, X., {Forveille}, T., {et~al.} 2003, in IAU Symposium
  211 ASP Conference Series, ed. E.~{Martin}, 311

\bibitem[{{Pier} {et~al.}(2003){Pier}, {Munn}, {Hindsley}, {Hennessy}, {Kent},
  {Lupton}, \& {Ivezi\'c}}]{Pier-2003:a}
{Pier}, J.~R., {Munn}, J.~A., {Hindsley}, R.~B., {et~al.} 2003, ApJ, 125, 1559

\bibitem[{{Pourbaix} {et~al.}(2003){Pourbaix}, {Platais}, {Detournay},
  {Jorissen}, {Knapp}, \& {Makarov}}]{Pourbaix-2003:b}
{Pourbaix}, D., {Platais}, I., {Detournay}, S., {et~al.} 2003, A\&A, 399, 1167

\bibitem[{{Richards} {et~al.}(2001){Richards}, {Fan}, {Schneider}, {Vanden
  Berk}, {Strauss}, {York}, {Anderson}, {Anderson}, {Annis}, {Bahcall},
  {Bernardi}, {Briggs}, {Brinkmann}, {Brunner}, {Burles}, {Carey}, {Castander},
  {Connolly}, {Crocker}, {Csabai}, {Doi}, {Finkbeiner}, {Friedman}, {Frieman},
  {Fukugita}, {Gunn}, {Hindsley}, {Ivezi{\' c}}, {Kent}, {Knapp}, {Lamb},
  {Leger}, {Long}, {Loveday}, {Lupton}, {McKay}, {Meiksin}, {Merrelli}, {Munn},
  {Newberg}, {Newcomb}, {Nichol}, {Owen}, {Pier}, {Pope}, {Richmond},
  {Rockosi}, {Schlegel}, {Siegmund}, {Smee}, {Snir}, {Stoughton}, {Stubbs},
  {SubbaRao}, {Szalay}, {Szokoly}, {Tremonti}, {Uomoto}, {Waddell}, {Yanny}, \&
  {Zheng}}]{Richards-2001:a}
{Richards}, G.~T., {Fan}, X., {Schneider}, D.~P., {et~al.} 2001, AJ, 121, 2308

\bibitem[{{Skrutskie}(1997)}]{Skrutskie-1997:a}
{Skrutskie}, M. 1997, S\&T, 94, 46

\bibitem[{{Smith} {et~al.}(2002){Smith}, {Tucker}, {Kent}, {Richmond},
  {Fukugita}, {Ichikawa}, {Ichikawa}, {Jorgensen}, {Uomoto}, {Gunn}, {Hamabe},
  {Watanabe}, {Tolea}, {Henden}, {Annis}, {Pier}, {McKay}, {Brinkmann}, {Chen},
  {Holtzman}, {Shimasaku}, \& {York}}]{Smith-2002:b}
{Smith}, J.~A., {Tucker}, D.~L., {Kent}, S., {et~al.} 2002, AJ, 123, 2121

\bibitem[{{Smol{\v c}i{\' c}} {et~al.}(2004){Smol{\v c}i{\' c}}, {Ivezi{\' c}},
  {Knapp}, {Lupton}, {Pavlovski}, {Illi{\' c}}, {Schlegel}, {Smith}, {McGehee},
  {Silvestri}, {Hawley}, {Rockosi}, {Gunn}, {Strauss}, {Fan}, {Eisenstein}, \&
  {Harris}}]{Smolcic-2004:a}
{Smol{\v c}i{\' c}}, V., {Ivezi{\' c}}, {\v Z}., {Knapp}, G.~R., {et~al.} 2004,
  ApJ, (submitted)

\bibitem[{{Wielen}(1996)}]{Wielen-1996}
{Wielen}, R. 1996, A\&A, 314, 679

\bibitem[{{York} {et~al.}(2000){York}, {Adelman}, {Anderson}, {Anderson},
  {Annis}, {Bahcall}, {Bakken}, {Barkhouser}, {Bastian}, {Berman}, {Boroski},
  {Bracker}, {Briegel}, {Briggs}, {Brinkmann}, {Brunner}, {Burles}, {Carey},
  {Carr}, {Castander}, {Chen}, {Colestock}, {Connolly}, {Crocker}, {Csabai},
  {Czarapata}, {Davis}, {Doi}, {Dombeck}, {Eisenstein}, {Ellman}, {Elms},
  {Evans}, {Fan}, {Federwitz}, {Fiscelli}, {Friedman}, {Frieman}, {Fukugita},
  {Gillespie}, {Gunn}, {Gurbani}, {de Haas}, {Haldeman}, {Harris}, {Hayes},
  {Heckman}, {Hennessy}, {Hindsley}, {Holm}, {Holmgren}, {Huang}, {Hull},
  {Husby}, {Ichikawa}, {Ichikawa}, {Ivezi{\' c}}, {Kent}, {Kim}, {Kinney},
  {Klaene}, {Kleinman}, {Kleinman}, {Knapp}, {Korienek}, {Kron}, {Kunszt},
  {Lamb}, {Lee}, {Leger}, {Limmongkol}, {Lindenmeyer}, {Long}, {Loomis},
  {Loveday}, {Lucinio}, {Lupton}, {MacKinnon}, {Mannery}, {Mantsch}, {Margon},
  {McGehee}, {McKay}, {Meiksin}, {Merelli}, {Monet}, {Munn}, {Narayanan},
  {Nash}, {Neilsen}, {Neswold}, {Newberg}, {Nichol}, {Nicinski}, {Nonino},
  {Okada}, {Okamura}, {Ostriker}, {Owen}, {Pauls}, {Peoples}, {Peterson},
  {Petravick}, {Pier}, {Pope}, {Pordes}, {Prosapio}, {Rechenmacher}, {Quinn},
  {Richards}, {Richmond}, {Rivetta}, {Rockosi}, {Ruthmansdorfer}, {Sandford},
  {Schlegel}, {Schneider}, {Sekiguchi}, {Sergey}, {Shimasaku}, {Siegmund},
  {Smee}, {Smith}, {Snedden}, {Stone}, {Stoughton}, {Strauss}, {Stubbs},
  {SubbaRao}, {Szalay}, {Szapudi}, {Szokoly}, {Thakar}, {Tremonti}, {Tucker},
  {Uomoto}, {Vanden Berk}, {Vogeley}, {Waddell}, {Wang}, {Watanabe},
  {Weinberg}, {Yanny}, \& {Yasuda}}]{York-2000:a}
{York}, D.~G., {Adelman}, J., {Anderson}, J.~E., {et~al.} 2000, AJ, 120, 1579

\end{thebibliography}

\end{document}